\documentclass[conference]{IEEEtran}
\IEEEoverridecommandlockouts
\usepackage{cite}
\usepackage{amsmath,amssymb,amsfonts}

\usepackage{textcomp}
\usepackage{xcolor}
\def\BibTeX{{\rm B\kern-.05em{\sc i\kern-.025em b}\kern-.08em
    T\kern-.1667em\lower.7ex\hbox{E}\kern-.125emX}}

\usepackage[skins]{tcolorbox}

\newtcolorbox{myframe}[2][]{%
  enhanced,colback=white,colframe=black,coltitle=black,
  sharp corners,boxrule=0.4pt,
  fonttitle=\itshape,
  attach boxed title to top left={yshift=-0.3\baselineskip-0.4pt,xshift=2mm},
  boxed title style={tile,size=minimal,left=0.5mm,right=0.5mm,
    colback=white,before upper=\strut},
  title=#2,#1
}

\usepackage{booktabs}
\usepackage{array}
\usepackage{graphicx}
\usepackage{graphics}
\usepackage{caption}
\usepackage{balance}
\usepackage{enumitem}
 \usepackage[disable]{todonotes}
\definecolor{airforceblue}{rgb}{1.74, 0.4, 0.85}
 \definecolor{airforceblue}{HTML}{81D8D0}

\usepackage{pifont}

\definecolor{mypink1}{rgb}{0.858, 0.188, 0.478}

 \usepackage{multirow}
 \usepackage{hhline}
\usepackage[ruled, vlined, linesnumbered]{algorithm2e}
\SetAlFnt{\footnotesize}

\usepackage{setspace}
\usepackage{algpseudocode}
\algrenewcommand\algorithmicindent{0.45em}
\usepackage{float}
\usepackage{upgreek}

\newcommand{\para}[1]{\vspace{1mm}\noindent\textbf{#1.}}
 
\definecolor{commentgreen}{rgb}{0,0.5,0}

\newcommand{\parab}[1]{\vspace{1mm}\noindent\textbf{#1}}
 
\usepackage{algpseudocode}
\usepackage{float}
\usepackage{caption}

\usepackage{upgreek}

\usepackage{relsize}

\makeatletter
\newcommand{\thickhline}{%
	\noalign {\ifnum 0=`}\fi \hrule height 1pt
	\futurelet \reserved@a \@xhline
}

\usepackage{listings}
\algnewcommand\algorithmicinput{\textbf{Input:}}
\algnewcommand\INPUT{\item[\algorithmicinput]}
\algnewcommand\algorithmicoutput{\textbf{Output:}}
\algnewcommand\OUTPUT{\item[\algorithmicinput]}

\SetCommentSty{mycommfont}

\usepackage{tikz}
\newcommand*\circled[1]{\tikz[baseline=(char.base)]{
            \node[shape=circle,draw,inner sep=0.8pt] (char) {#1};}}

\algnewcommand\algorithmicforeach{\textbf{for\  each}}
\algdef{S}[FOR]{ForEach}[1]{\algorithmicforeach\ #1\ \algorithmicdo}

\usepackage{amsthm}
\usepackage{amsmath}
\usepackage[utf8]{inputenc}
\usepackage[english]{babel}

\theoremstyle{definition}
\newtheorem{definition}{Definition}[section]
  \newtheorem{exmp}{Example}[section]

\newcounter{Asterios}
\stepcounter{Asterios}

\definecolor{andrapink}{HTML}{b7306b}
\newcounter{Andra}
\stepcounter{Andra}

\newcounter{Ziyu}
\stepcounter{Ziyu}

\newcounter{Mingliang}
\stepcounter{Mingliang}

\newcounter{RihanNOC}
\stepcounter{RihanNOC}

\newcounter{Wenbo}
\stepcounter{Wenbo}

\usepackage[%
font={bf},
labelfont=bf,
format=hang,
]{caption}

\usepackage{subcaption}
\usepackage{soul}

\begin{document}

\title{Amalur: Data Integration Meets Machine Learning}

\author{
    \IEEEauthorblockN{
     Rihan Hai \qquad  Christos Koutras \qquad Andra Ionescu \qquad Ziyu Li \qquad Wenbo Sun \\ \qquad Jessie van Schijndel \qquad Yan Kang$^*$  \qquad Asterios Katsifodimos}
    \IEEEauthorblockA{Delft University of Technology, $^*$WeBank}
    \IEEEauthorblockA{\{initial.lastname\}@tudelft.nl, yangkang@webank.com}
}

\maketitle


\begin{abstract}
Machine learning (ML) training data is often scattered across disparate collections of datasets, called \emph{data silos}. This fragmentation poses a major challenge for data-intensive ML applications: integrating and transforming data residing in different sources demand a lot of manual work and computational resources. 
With data privacy and security constraints, data often cannot leave the premises of data silos,  
hence model training should proceed in a decentralized manner.
In this work, we present a vision of how to bridge the traditional data integration (DI) techniques with the requirements of modern machine learning. 
We explore the possibilities of utilizing metadata obtained from data integration processes for improving the effectiveness and efficiency of ML models. Towards this direction, we analyze two common use cases over data silos, feature augmentation and federated learning. Bringing data integration and machine learning together, we highlight new research opportunities from the aspects of systems,  representations, factorized learning and federated learning.
\end{abstract}

\section{Introduction}
\label{sec:intro}

The accuracy of an ML model heavily depends on the  training data. 
In real world applications, often the  data is not stored in a central database or file system, but  spread  over different  data silos. 
Take, for instance,  drug risk prediction: the features can reside in datasets collected from clinics, hospitals, pharmacies, and laboratories distributed geographically \cite{bos2018prediction}. 
Another example is 
training  models for keyboard stroke prediction: training requires data from  millions of phones \cite{hard2018federated}.

Data integration systems enable interoperability among multiple, heterogeneous sources, and provide a unified view for users. Notably, they allow us to describe data sources and their relationships \cite{doan2012principles}: $i)$ mappings  between different source  schemata, i.e., 
schema matching and mapping \cite{rahm2001survey, fagin2009clio} and $ii)$ linkages between data instances, i.e., data matching (also known as record linkage or entity resolution)  \cite{brizan2006survey}. 
Yet,  a data integration system's goal is to facilitate   query answering or data transformation  over silos, and not to directly support machine learning applications. As a result, practitioners nowadays tackle silos with DI systems and ML tooling separately, as shown in the following.

\para{Running example} Consider the feature augmentation example in Figure~\ref{fig:motiv}, where the downstream ML task is to predict the mortality (binary classification) of patients based on information scattered across tables maintained by different departments in the same hospital. Data from the ER department are stored in a base table \textit{$S_1$(m,n,a,hr)}, which has the label column $m\ (mortality) $, and feature columns $a\ (age)$ and $hr\ (resting\ heart\ rate)$. To improve the model's accuracy, a data discovery system is employed to discover a related table \textit{$S_2$(m,n,a,o,dd)} (Figure~\ref{fig:motiv}b), with information coming from the pulmonary department. This table brings information about a new feature column $o\ (oxygen)$, which shows the blood oxygen level.
The label column and the selected feature columns constitute  the schema of the table for downstream ML models, i.e., \textit{T(m, a, hr, o)}, which we refer to as the \textit{target table schema} or \emph{mediated schema}.

\begin{figure}[t]
\centering
\includegraphics[width=0.30\textwidth]{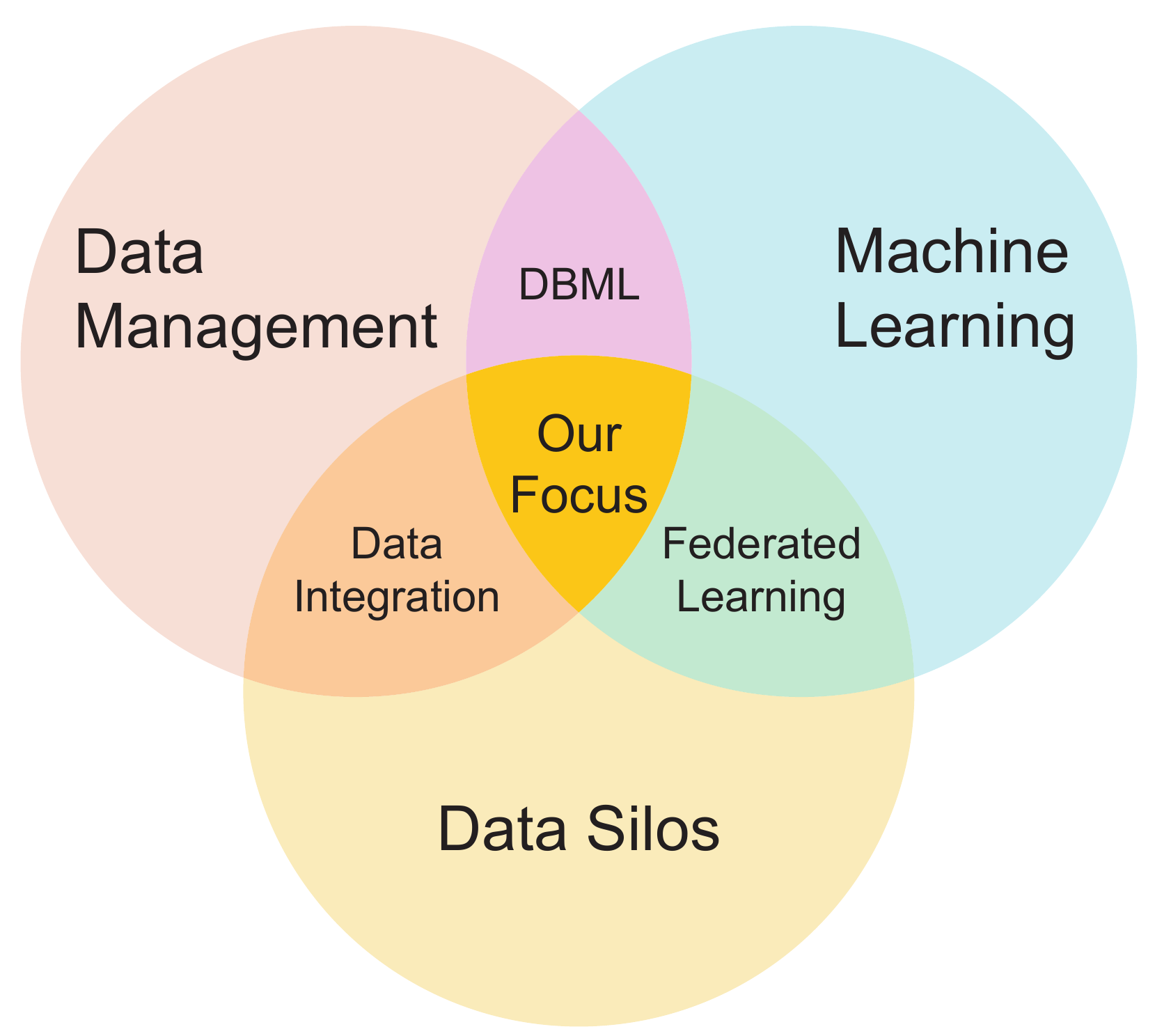}
\caption{Scope of this line of work}
\label{fig:venn}
\vspace{-0.3cm}
\end{figure}

\para{Data integration, data management and ML}  Figure \ref{fig:venn} illustrates our problem scope.  
Recent advances of \emph{in-database machine learning} \cite{Makrynioti2019survey, zhou2020database, 10.1007/978-3-030-35514-2_32}, mainly consider a single database instead of data silos\footnote{The intersection of  data management and ML (DBML) is two-fold: ML for DB, and DB for ML. Machine learning has been applied to improve key operations of data integration such as schema matching \cite{ DBLP:journals/tois/AlserafiARC20, 10.1145/3318464.3389742}, and data matching \cite{kopcke2010evaluation, das2017falcon, 9338287, 10.1145/3318464.3389742}. In this paper we focus on data management for machine learning.  Except for data cleaning \cite{krishnan2016activeclean, krishnan2017boostclean, 9458702}, little has been discussed in terms of using the key DI operations to facilitate machine learning \cite{dong2018data}.}.  
Traditional data integration solves the data management issues of data silos \cite{DBLP:conf/vldb/HalevyRO06}. In a similar way, federated learning (FL) studies machine learning with training data residing in  data silos~\cite{yang2019federated}. In this paper, we argue that when data management, data silos, and machine learning meet, there is a new set of challenges and opportunities for research and optimization.

\para{Issues with the separation of DI and ML} As shown in Figure~\ref{fig:motiv}c-d, to use the data from the two tables $S_1$ and $S_2$, a data scientist would need to rely on a data integration system, or else manually find the schema mapping and entity resolution between the two given tables. We elaborate on the explanation of schema mappings in Section~\ref{ssec:mapping}.
Then, a data integration system can integrate these source tables by merging the mapped columns and   linked entities (i.e., matched rows).
Finally, it materializes the data instances of the target  table $T$ and exports it to downstream ML applications. Such a process usually involves massive manual work and computation overhead, e.g., joining tables. Meanwhile, it assumes that users are aware of the data sources, know the principles of data integration, and/or are familiar with DI tools. 
In other words, there is great potential to utilize DI techniques to  reduce the human burden and automate ML pipelines.
\todo[]{add no-materialize}

\para{Research vision \& question} 
Data integration is a well-studied research area with mature logic-based theoretical frameworks, techniques, and systems \cite{lenzerini2002data, DBLP:conf/vldb/HalevyRO06, DBLP:conf/pods/GolshanHMT17}. 
We  envision novel systems that combine DI techniques (schema matching, schema mapping, entity resolution, query reformulation, etc.) and ML pipelines (e.g., feature selection and augmentation, model training), 
while also expanding to more ML philosophies (e.g., federated learning). As the starting point,   we ask a  fundamental question:

\vspace{2mm}
\noindent \textit{$Q$: Can we use data integration metadata to improve the effectiveness and efficiency of ML model training?}
\vspace{2mm}

With this line of work, we aim   to investigate  whether the metadata obtained from data integration, specifically the output of \emph{schema matching} and \emph{entity resolution}, can benefit downstream ML tasks. To this end, we present new research challenges that arise when we design a novel data integration system, which extends the concepts of query rewriting and data transformations for the needs of ML model training, and saves costs associated to intermediate result materialization and the exporting of target tables.

In this paper we focus on these challenges in four aspects: \emph{system-design}, \emph{metadata representation}, \emph{performance}, and \emph{privacy}. The contributions of this paper go as follows:
\begin{itemize}[leftmargin=*]
\item \textit{{System-design (Section~\ref{sec:sys}).}} We demonstrate the design of \emph{Amalur}, a novel data integration system, which  supports end-to-end, scalable machine learning pipelines over data silos. We elaborate on the research challenges of building such a system.

\item \textit{{Metadata representations (Section~\ref{sec:rewrite}).}} 
We propose   matrix-based representations for data integration metadata, which  capture   $i)$ column matches, $ii)$ row matches, and $iii)$ redundancies between data sources and the target table. We also discuss and compare other available alternatives as representations for DI metadata.

\begin{figure}[t]
\centering
\includegraphics[width=\columnwidth]{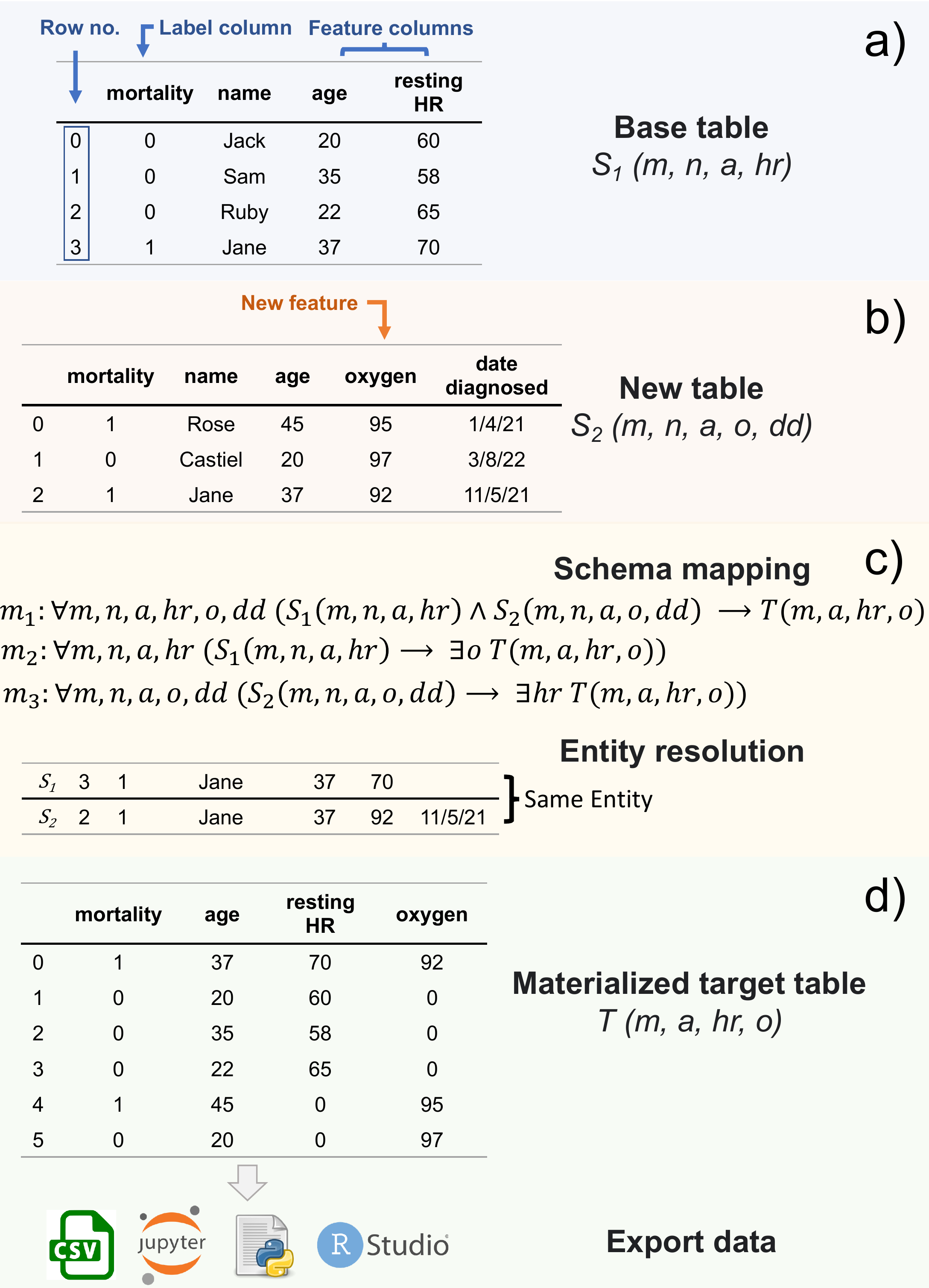}
\caption{Traditional integration of data silos for ML}
\label{fig:motiv}
\vspace{-0.3cm}
\end{figure}

\item \textit{{Performance. (Section~\ref{sec:opp})}} 
We highlight the new opportunities for utilizing DI metadata to improve the time-wise efficiency of  ML model training over data silos.  

\item \textit{{Privacy (Section~\ref{sec:fl}).}}
We discuss the research challenges of improving vertical federated learning  with data integration metadata. 

\end{itemize}

\section{Amalur: An ML-oriented data integration system}
\label{sec:sys}

In this section, we introduce our proposed system Amalur. We  explain the challenges of two common ML use cases, namely feature augmentation and federated learning, and discuss how Amalur can tackle these challenges.

 \begin{figure}[tb]
\centering
\includegraphics[width=\linewidth]{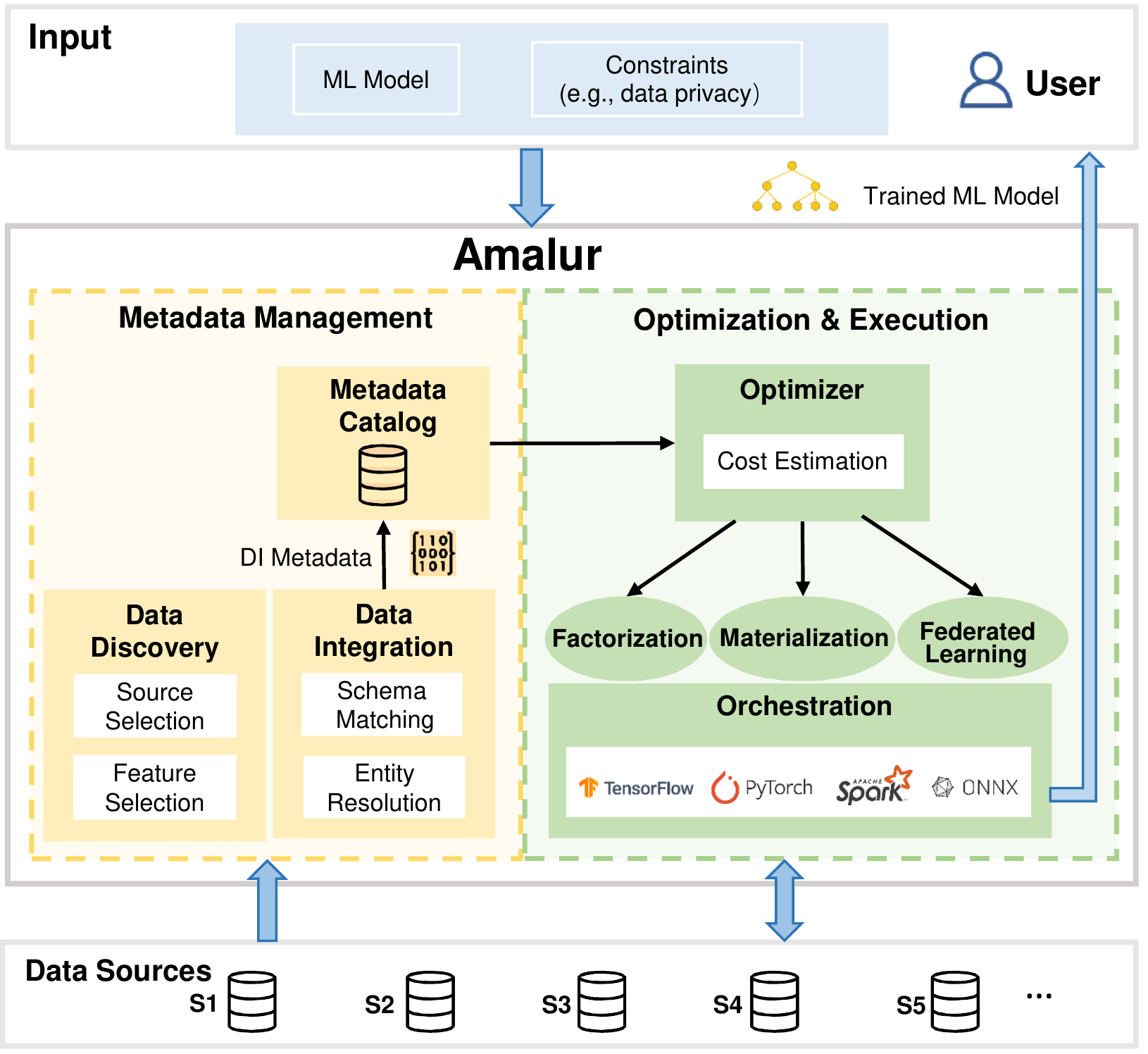}
 \vspace{-0.2cm}
\caption{An overview of   Amalur }
\label{fig:DDSP}
\vspace{-0.4cm}
\end{figure}


\subsection{Amalur overview}
We are currently developing \emph{Amalur}, a \emph{model lake} system that is based on our work on data lakes \cite{hai2016constance} and model zoos \cite{li2022metadata}.
With DI metadata, Amalur features solving the challenges of scalable training of ML models over data silos, reducing the manual work of integrating the data and speeding up model training. 
Figure~\ref{fig:DDSP} provides a high-level overview of Amalur with key components relevant to this paper.

\para{User inputs}
Amalur allows users (e.g., physicians, data scientists) to train models on data silos. The user may already have an ML model, e.g., defined in Python scripts.
There might also be  constraints specific to a user and silos, e.g., data privacy  regulations such as GDPR \cite{voigt2017eu}. 


\para{Hybrid metadata catalog} 
One of the fundamental components of Amalur is the metadata catalog. It stores the metadata of data 
 and ML models. 
Data-related metadata includes the basic metadata and   data integration metadata. 
Collected from the silos, the basic metadata describes
 each data source, e.g., source table schema, data types, integrity constraints, data provenance information such as silo location. In this work, by \emph{data integration metadata} we refer to the metadata generated during the data integration process, e.g., column relationships from schema matching and row matching from entity resolution. 
Model-related metadata include the model execution environment, configurations (e.g., hyper-parameters), input/output (e.g., prediction class), evaluation performance (e.g., model accuracy), etc. The metadata catalog also keeps track of the connections between the model and its training datasets.   
We have addressed the representations of basic metadata of source tables \cite{quix2016metadata} and ML models \cite{li2022metadata}. In this work, we focus on data integration metadata and explain their matrix-based representation in Section~\ref{sec:rewrite}.

\para{Optimization and coordination} Another essential component is  the optimizer.
Given the input ML model, constraints and metadata from the metadata catalog, the optimizer estimates the cost and decides how the ML model training will be performed over silos: 1) the ML model is decomposed and the computation is pushed down to the source tables stored in the silos, i.e., factorization;
2) the source tables are joined and the generated target table is exported for model training, i.e., materialization;
3) the learning process is split given  privacy-preservation constraints, i.e., federated learning. We  elaborate on the details in Section~\ref{sec:opp} and Section~\ref{sec:fl}. 

\para{Compilation, orchestration and distribution} 
Finally, the execution plan from the optimizer is compiled into concrete programs according to the execution environment (e.g., TensorFlow, PyTorch, Spark, ONNX). The compiled binaries are executed either in a central orchestrator or multiple remote orchestrators. Specifically, for factorized executions, the executables are shipped to the different data silos and return results 
needed in a centralized computation.




\subsection{Silos problem: ML use cases and DI formulation}
\label{ssec:4cases}

\begin{table*}[tb]
\centering
\small
\caption{Four example data integration scenarios for feature augmentation and federated learning}
\label{tbl:4cases}
\resizebox{\textwidth}{!}{
\begin{tabular}{lp{55pt}ll} 
\hline
\specialrule{0em}{1pt}{1pt}
\multirow{2}{*}{\textbf{No.}} & \textbf{Dataset}                         & \multirow{2}{*}{\textbf{Schema mappings}}                                           &  \multirow{2}{*}{\textbf{Example use cases}}\\
& \textbf{Relationship} &  & \\
\hline \hline
\specialrule{0em}{1pt}{1pt}
1   & Full
outer join                                & \begin{tabular}[c]{@{}l@{}} $ m_1:\ \forall m,n,a,hr,o,dd \ (S_1(m,n,a,hr) \wedge S_2(m,n,a,o,dd)   \rightarrow T(m,a,hr,o) )$\\$m_2:\ \forall m,n,a,hr \  (S_1(m,n,a,hr) \rightarrow  \exists o \ T(m,a,hr,o)) $\\$m_3:\ \forall m,n,a,o,dd \  (S_2(m,n,a,o,dd) \rightarrow  \exists hr \ T(m,a,hr,o)) $\end{tabular}  
& \begin{tabular}[c]{@{}l@{}}Feature augmentation,\\ Federated learning, \ldots \end{tabular}\\ 
\hline
\specialrule{0em}{1pt}{1pt}
2   & \textcolor[rgb]{0.251,0.251,0.251}{Inner join} &  $ m_1:\ \forall m,n,a,hr,o,dd \ (S_1(m,n,a,hr) \wedge S_2(m,n,a,o,dd)   \rightarrow T(m,a,hr,o) )$            & \begin{tabular}[c]{@{}l@{}}Feature augmentation,\\ (Vertical) federated learning, \ldots\end{tabular}                                          \\ 
\hline
\specialrule{0em}{1pt}{1pt}
3   & Left
join                                      & \begin{tabular}[c]{@{}l@{}} $ m_1:\ \forall m,n,a,hr,o,dd \ (S_1(m,n,a,hr) \wedge S_2(n,a,o,dd)   \rightarrow T(m,a,hr,o) )$\\$m_2:\ \forall m,n,a,hr \  (S_1(m,n,a,hr) \rightarrow  \exists o \ T(m,a,hr,o)) $\end{tabular}    
& \begin{tabular}[c]{@{}l@{}}Feature augmentation,\\  (Vertical) federated learning, \ldots\end{tabular}\\ 
\hline
\specialrule{0em}{1pt}{1pt}
4   & Union
& \begin{tabular}[c]{@{}l@{}}$m_2:\ \forall m,n,a,hr,o \  (S_1(m,n,a,hr,o) \rightarrow   T(m,a,hr,o)) $\\$m_3:\ \forall m,n,a,hr,o,dd \  (S_2(m,n,a,hr,o,dd) \rightarrow   T(m,a,hr,o)) $\end{tabular}     & \begin{tabular}[c]{@{}l@{}}Data sample augmentation,\\ (Horizontal) federated learning, \ldots\end{tabular}   \\
\hline
\end{tabular}
}
\vspace{-0.3cm}
\end{table*}

In Table~\ref{tbl:4cases}, we provide two representative  ML use cases where training data could come from silos, i.e., feature augmentation and federated learning. 
Existing solutions for factorization over joins 
\cite{chen2017towards, 10.1145/3299869.3319878, cheng_efficient_2021} mostly tackle inner joins. In this work, we also deal with left joins, full outer joins, and unions. 
As shown in  Table~\ref{tbl:4cases}, for the use cases of feature augmentation and federated learning (and possibly many more), 
the dataset relationships  between source tables and desired target table,  can be captured by a class of well-studied data dependencies, i.e.,  tuple-generating dependencies (tgds) \cite{beeri1984proof, Fagin2009}, which are the commonly used   formalism in  data integration studies. 


\parab{Use case 1: Feature augmentation}  is the exploratory process of finding new datasets and
selecting features that help improve the ML model performance \cite{chepurko2020arda, esmailoghli2021cocoa, kumar2016join}. Figure~\ref{fig:motiv}b shows an example: starting from a base table $S_1$, we augment the features by introducing the  table $S_2$ and selecting the new feature $o\ (oxygen)$.

\noindent\textbf{Use case 2: Federated learning} ~\cite{yang2019federated}
studies how to build joint ML models over data silos (e.g., enterprise data warehouses, edge devices) without compromising privacy, which follows a decentralized learning paradigm. 
Similar to the problem setting of virtual data integration \cite{doan2012principles}, FL assumes that the source data is not collected and stored at a central data store but stays at the local data stores. 
According to how the feature space and sample space are partitioned among the data sources, FL can be  categorized as vertical federated learning (VFL) and horizontal federated learning (HFL) \cite{yang@2019fml}. 
For VFL, data sources share the overlapping data instances, but the feature columns partially overlap or not. For HFL, data sources share the overlapping feature columns, while the data instances may overlap.

\parab{Example 1 (full outer join)} is explained  for feature augmentation in Figure~\ref{fig:motiv} and Example~\ref{exmp:map_tgd}.
It can also be seen as a general case of federated learning, where  sources have similar schemas, and data instances (entities) that may, or may not overlap with each other.

\parab{Example 2 (inner join)} represents the DI scenario where only overlapped rows in two sources will be transformed, i.e., an inner join between $S_1$ and $S_2$ followed by a projection on columns $m, a, hr, o$.
It can be used to describe the feature augmentation processes where fewer missing values are preferred. Such dataset relationships also reside in a VFL use case, where data sources share the sample space (overlapped rows) but not necessarily the feature space (overlapped feature columns).  

\parab{Example 3 (left join)}  shows a left join between $S_1$ and $S_2$. 
Compared to Example 1, we  slightly change the schema of $S_2$ by dropping the label column $m$.  Example 3  describes another typical feature augmentation scenario for supervised learning: only the base table $S_1$ contains the label column. Thus, when adding features from the new table $S_2$, only rows overlapped with $S_1$ will be selected.
Example 3 can be used to describe the VFL cases where not all but specific sources hold the labels for supervised model training.

\parab{Example 4 (union)} is a special case of Example 1, where  $S_1$ and $S_2$ do not share any rows. We modify the schemas of $S_1$ and $S_2$ such that they share the same set of feature columns which are mapped to the target schema $T$. 
Example 4 can represent the scenario when a new table is selected to bring more data samples. 
Alternatively, it can describe the HFL scenario where data sources share feature columns but not data samples.

\subsection{Amalur workflows for the two ML use cases }
\label{ssec:explain use cases}

\para{Amalur workflow for feature augmentation} When there is no privacy constraint,  Amalur determines the computation mechanism based on cost estimation. If the computation is performed in a factorized manner, the model is decomposed and pushed down to silos. 
If the computation is performed in a materialized manner, Amalur will integrate the source datasets and generate the target table. 

\para{Amalur workflow for federated learning}
In the existence of privacy constraints, Amalur will conduct privacy-preserving data integration operations over the silos \cite{DBLP:conf/sigmod/ScannapiecoFBE07}, and split the learning process over the silos.
The central orchestrator will coordinate communication between silos, and the encryption/decryption during aggregating the results and updating the   weights.

\section{Representation: A Tale of Three Matrices}
 \label{sec:rewrite}
 
In this section,  we discuss the  representations  that capture the metadata of data integration, which enable optimizing ML tasks. 
When combining the functions of data integration and machine learning in one system such as Amalur, one core task is how to represent the DI metadata. By DI metadata, we refer to the information that describes \emph{i}) the selected data sources, e.g., table schemas and the number of rows, and the number of selected sources; \emph{ii}) the relevance and overlap between data sources, e.g., column matching between source tables (schema matching), schema-level correspondences between source tables and the target table (schema mapping), and row matching between source tables (entity resolution). 
The challenge is that \textit{the representation needs to be expressive enough to capture the DI metadata, while bringing little overhead for model training}.

We define three logical-level representations: \emph{mapping matrix} for preserving the column mapping (Section~\ref{ssec:mm_gen}), 
 \emph{indicator matrix} for row matching (Section~\ref{sssec:imGen}), and   \emph{redundancy matrix} for data redundancy (Section~\ref{ssec:RM}). We choose  matrix-based representations, as they facilitate direct computation with linear algebras without the need of additional transformation, which we illustrate in Section~\ref{sec:opp}. Finally, 
in Section~\ref{ssec:dis}, we inspect the implementation of such matrix-based representations at the physical level. 


 \begin{table}[t]
\small
\begin{center}	
\caption{Notations used in the paper}
\vspace{-0.1cm}
\label{tbl:nota}
\begin{tabular}{|l|l|}
\hline
\textbf{Notation}  & \textbf{Description}                   \\ \hline
$T$/$S_k$     & Target table/the k-th source table    \\ \hline
$D_k$     & Processed   k-th source table in matrix form     \\ \hline
$c_T/c_{S_k}$     & Number of mapped columns in $T/S_k$ \\ \hline
$r_{T}$/$r_{S_k}$   & Number of mapped rows in $T$/$S_k$        \\ \hline
$M_k/CM_k$ & Full/compressed mapping matrix for  $S_k$       \\ \hline
$I_k/CI_k$ & Full/compressed indicator matrix for  $S_k$       \\ \hline
$R_k$ & Redundancy matrix for  $S_k$       \\ \hline
\end{tabular}
\end{center}
  \vspace{-0.4cm}
\end{table}

\begin{figure*}[tb]
\centering
\includegraphics[width=\textwidth]{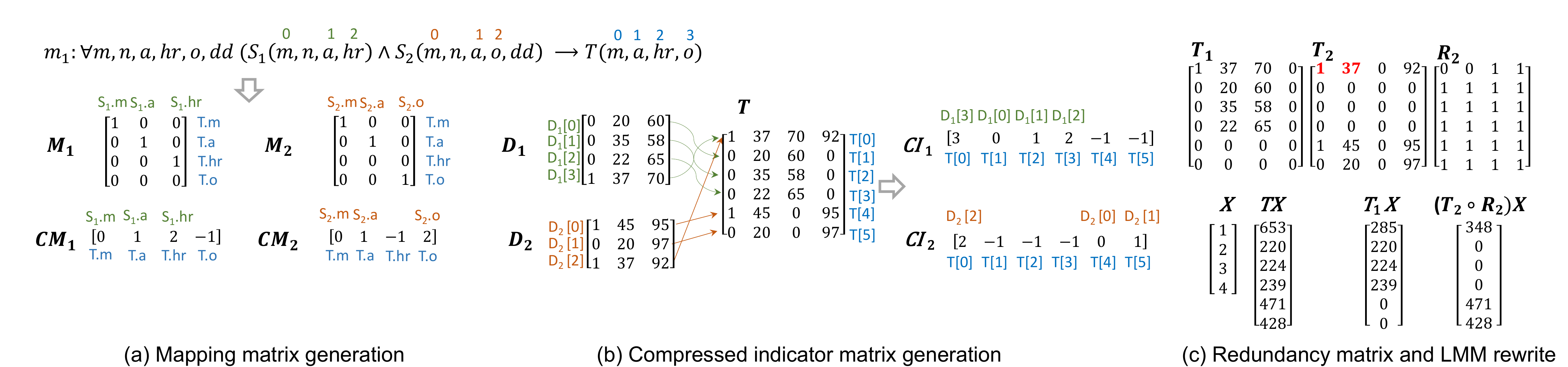}
\caption{Mapping, indicator, and redundancy matrices of the  running example}
\label{fig:allMs}
\end{figure*}

\subsection{Mapping matrix}
\label{ssec:mm_gen}

\para{Preliminaries}
\label{ssec:mapping}
\emph{Schema mappings}  lay at the heart of data integration and  data exchange. Let $\mathbf{S}$ and $\mathbf{T}$ be a source relational schema and a target relational schema sharing no relation
symbols.
A \emph{schema mapping} $\mathcal{M}$ between $\mathbf{S}$ 
and $\mathbf{T}$ is
a triple $\mathcal{M} = \left \langle\mathbf{S}, \mathbf{T}, \Sigma \right \rangle$, 
where $\Sigma$ is a set of dependencies over $ \left \langle \mathbf{S},\mathbf{T}\right \rangle$.
The dependencies $\Sigma$ can be 
expressed as logical
formulas  over source and target schemas. One of the most commonly used mapping languages is \emph{source-to-target tuple generating dependencies (s-t tgd)}  \cite{beeri1984proof, Fagin2009}, which are also known as Global-Local-as-View (GLAV) assertions 
\cite{lenzerini2002data}.  
An s-t tgd is a first-order sentence in the form of  
$\forall \mathbf{x} \ (\varphi (\mathbf{x}) \rightarrow \exists \mathbf{y} \ \psi
(\mathbf{x},\mathbf{y} ))$, where $\varphi (\mathbf{x})$ is a conjunction of atomic formulas over the source schema $\mathbf{S}$, and $\psi
(\mathbf{x},\mathbf{y} )$  is a conjunction of atoms over the target schema $\mathbf{T}$.
\vspace{-0.2cm}
\begin{exmp} 
	\label{exmp:map_tgd} In Figure~\ref{fig:motiv}c, 
$m_1$, $m_2$, and $m_3$ are all tgds.
We represent mapped attributes  with the same variable names, e.g., $S_1.m$ and $S_2.m$.
The tgd $m_1$  specifies that the overlapped rows of $S_1$ and $S_2$ are added to $T$ ($\wedge$ denotes a natural join between $S_1$ and $S_2$); $m_2$  and  $m_3$ indicate that all rows of $S_1$ and $S_2$ will be transformed to generate new tuples in $T$, respectively. Among the three tgds, it is the union relationship. 
The three tgds together, describe that the instances in $T$ are obtained via a full outer join between the datasets $S_1$ and $S_2$.
\end{exmp}
\vspace{-0.2cm}	


\para{Gaps} Tgds are first-order sentences specifying schema mappings. A DI system often generates schema mappings as executable data transformation scripts, e.g., SQL, which transform the source data instances and materialize the target table $T$.
In contrast, the fundamental language of ML models    is linear algebra. To embed schema mappings in an end-to-end ML pipeline, we need a novel representation for schema mappings, which is  compatible with algebraic computation in ML model training.

\para{Matrix-based representation for schema mappings}
Schema mappings contain the information about the mapped columns between source and target tables. We   define the \emph{mapping matrix} to preserve  such   column mappings.
As a preparation step, we add  ID numbers to  mapped columns as shown in Figure~\ref{fig:allMs}a. In Table~\ref{tbl:nota} we summarize the notations used in this paper. We abuse the notation a little and refer to both a target/source table name and its schema with  $T/S_k$. 


\begin{definition}[\textit{Mapping matrix}] Mapping matrices between source tables $S_1, S_2, ..., S_n$ and target table $T$ are a set of binary matrices $\mathbf{M}= \{M_1, ..., M_n\}$. $M_k \ (k\in [1, n])$ is a matrix with the shape $c_{T} \times c_{S_k}$, where  

\vspace{-0.2cm}
	\small{
	\begin{align*}
 		\begin{split} 
 		M_k[i,j] =  \begin{cases}1, &if\ j^{th}\ column\  of\  S_k\  is\  mapped\  to\ \\  & the\    i^{th}\ column\  of\  T\\0, & otherwise\end{cases}
 		\end{split}
\end{align*}
}
\label{def:mm}
\vspace{-0.4cm}
\end{definition}
Intuitively, in $M_k[i,j]$ the vertical coordinate $i$ represents the target table column while the horizontal coordinate $j$ represents the mapped source table column.  A value of $1$ in $M_k$ specifies the existence of  column correspondences between  $S_k$ and $T$, while the value $0$ shows that the current target table attribute has no corresponding column in $S_k$. Figure~\ref{fig:allMs}a shows the mapping matrices $M_1$ for $S_1$, and $M_2$ for $S_2$ of the running example. 

It is easy to see that the binary  mapping matrices are often sparse.
Because each attribute in the   source table $S_k$ is mapped to only one  attribute in $T$. 
Thus, in each row of $M_k$ at most one element is 1, while the rest are 0.  
Moreover, if an attribute of T does not have a mapped attribute in $S_k$, the corresponding row of the mapping matrix has only values of 0. For example, $ T.o$   (column ID: 3) does not have a mapped column in $S_1$, thus, the last row of  $M_1$ has only zeros, i.e., $M_1[3] = [0,0,0]$.
To solve the sparsity problem we apply a more compressed form of mapping matrices as follows.
\begin{definition}[\textit{Compressed mapping matrix}] Compressed mapping matrices between source tables $S_1, S_2, ..., S_n$ and target table $T$ are a set of row vectors $\mathbf{CM}= \{CM_1, ..., CM_n\}$. $CM_k \ (k\in [1, n])$ is a  row vector  of size $c_{T}$, where 

\vspace{-0.2cm}
	\small{
	\begin{align*}
 		\begin{split} 
 		CM_k[i]=    \begin{cases}j,&if\ j^{th}\ column\  of\  S_k\  is\  mapped\ \ to\ \\
 		& the\    i^{th}\ column\  of\  T\\-1,& otherwise\end{cases}
 		\end{split}
\end{align*}
}
\label{def:cmm}
\vspace{-0.2cm}
\end{definition}
We continue with the running example. Figure~\ref{fig:allMs}a illustrates the compressed mapping matrices $\mathbf{CM} = \{CM_1, CM_2\}$. They 
can be directly generated from schema mappings without the generation of the  mapping matrices $\mathbf{M} = \{M_1, M_2\}$. 
%



\subsection{Indicator matrix }
\label{sssec:imGen}

We use the \emph{indicator matrix} \cite{chen2017towards} (denoted as $I_k$) to preserve the row matching between each source table $S_k$ and the target table $T$. Similar to the mapping matrix, a binary indicator matrix could be very sparse and   its compressed form is preferred. 
Due to space restriction, we directly define the \emph{compressed indicator matrix}.

\begin{definition}[\textit{Compressed indicator matrix}] Compressed indicator matrices between source tables $S_1, S_2, ..., S_n$ and target table $T$ are a set of row vectors $\mathbf{CI}= \{CI_1, ..., CI_n\}$. $CI_k \ (k\in [1, n])$ is a row vector  of size $r_{T}$, where 
\vspace{-0.3cm}

	\small{
		\begin{align*}
		\vspace{-0.2cm}
			\begin{split} 
					CI_k[i] =  \begin{cases}j, &if\ the \ j^{th}\ row\  of\ S_k \   is\  mapped\  to\  \\
					& the\ i^{th}\ row\  of\  T\
				\\-1, & otherwise\end{cases}
			\end{split}
		\end{align*}
	}
	\label{def:cim}
	\vspace{-0.3cm}
\end{definition}


\vspace{-0.1cm}

Notably, for the downstream ML algorithms, not all but only partial data of an original source table will  participate in the computation as features or labels. 
Thus, we transform the original tables $S_1$ and $S_2$ in Figure~\ref{fig:motiv}a-b to their matrix forms $D_1$ and $D_2$  in Figure~\ref{fig:allMs}b, which only include the mapped columns.
Figure~\ref{fig:allMs}b shows the row matching of the running example and the compressed indicator matrices, $CI_1$ for $S_1$ and $CI_2$ for $S_2$.

\subsection{Redundancy matrix}
\label{ssec:RM}
Data integration systems often need to handle data redundancy when  multiple sources have overlapping values. Consider the  example in Figure~\ref{fig:motiv}, when a user query asks how many patients aged above 30 are in $S_1$ and $S_2$, the correct answer is three instead of four. That is, the overlapped row of \emph{Jane} should be counted only once. 
Such redundancy resides in 
the projection of shared rows on the overlapped columns. 
Similarly, to support ML models we also need to detect redundancy to avoid repeated computation, which might lead to erroneous results. Thus, we propose a declarative representation to capture   redundancy, i.e., \emph{redundancy matrix}.  
To prepare for its definition, we first discuss how each source table contributes to the target table materialization. 
With the mapping matrix $M_k$ and indicator matrix $I_k$, we can transform a source table $D_k$ to an intermediate matrix with the same shape as  $T$, denoted as $T_k$. 
\begin{equation*}
	\vspace{-0.1cm}
T_k =  I_kD_kM^T_k
\end{equation*}
Figure~\ref{fig:allMs}c shows $T_1$ and $T_2$ of the running example. The red values in $T_2$ are the repeated values that already appeared in $T_1$. 
It is easy to see that 
$T_k$ indicates the contribution from each source $S_k$.
However,  due to the aforementioned redundancy issue (red values in $T_2$), we cannot make a simple matrix addition to obtain the target table. 
For instance,  $T_1 + T_2 \neq T$ in Figure~\ref{fig:allMs}b-c.
This is why we need the redundancy matrix, which is defined below.
\begin{definition}[\textit{Redundancy matrix}] A redundancy matrix $R_k$ of source table $S_k$ is a binary matrix with the shape  of $r_T \times c_T$, where  

\vspace{-0.2cm}
	\small{
	\begin{align*}
 		\begin{split} 
 		R_k[i,j] =  \begin{cases}0, &if\ T_k[i,j]\  is\  redundant \\1, & otherwise\end{cases}
 		\end{split}
\end{align*}
}
\label{def:rm}
\vspace{-0.4cm}
\end{definition}
Note that before we say the data of a  source table is  redundant, we first need to specify which source table is the base table. 
For instance, in Example 1-3 of Table~\ref{tbl:4cases}, if we specify $S_1$ as the base table, then we consider the overlapped values in $S_2$ are redundant, and only need to generate a redundancy matrix for $S_2$. For completeness we can consider that the redundancy matrix for the base table is an \emph{all-ones matrix}, which has all the element  values equal to one. 
Figure~\ref{fig:allMs}c shows $R_2$ for $S_2$ given  the running example, which is computed based on mapping and indicator matrices of $S_1$ and $S_2$.

\subsection{Metadata representation as tensors}
\label{ssec:dis}




 The three proposed types of matrices offer a novel perspective on the data processing pipeline, where we can describe data, and data integration processes with linear algebra. Aside from the intuitive 2-dimensional matrix representation, we can use \emph{high-dimensional tensors} to integrate data and metadata. For example, the data matrix $D_k$ can be expanded to a third dimension where $M_k$ and $I_k$ adhere along values and primary keys respectively. As such, we can represent the data and data integration metadata with a more expressive data structure compatible with tensor algebra and recent advances in data processing \cite{xla,trt,pytorch, tqp,tcudb,tensordb,hummingbird}. 

As the fundamental language of machine learning, tensor algebra and the benefits it brings in efficiency \cite{xla,trt,pytorch} have attracted the considerable attention of the ML and DB research communities. Many recent works \cite{tqp,tcudb,tensordb,hummingbird} have started considering the integration of data processing and ML pipeline in unified tensor runtimes. Such a combination enables cross-optimizations between data processing and ML pipelines, a vital part of the Amalur system.

Furthermore, as dedicated tensor processing modules keep emerging, the tensor representation exhibits compatibility with new hardware. Google TPUs \cite{tpu} and recent Nvidia GPUs \cite{tensorcore} have built-in tensor cores where matrix multiplication can be computed in one single clock cycle, improving parallelism from the array-level of SIMD instructions to the matrix-level. The High Performance Computing community \cite{matrixparallel,nature} has started optimizing numeric algorithms for tensor cores. It is foreseen that the tensor-represented data processing can be significantly accelerated with co-evolving tensor algorithms, runtime, and dedicated tensor computing hardware.


\begin{myframe}{~Summary \& Opportunities~}
\emph{ --    DI metadata can be preserved in matrix representation.}
\\ \emph{ -- Representing DI metadata as tensors offers cross-optimization opportunities between DI and ML, and can help us leverage emerging tensor processing methods and hardware for speedup.}
\end{myframe}

\section{Algebraic computation over silos}
\label{sec:opp}
In this section, we dive into the research opportunities of conducting arithmetic computations over silos with data integration metadata. 
In the following, we first discuss 
the new challenges of generalizing the existing factorization techniques from a single database to data silos (Section~\ref{ssec:LAR}), and     cost estimation for choosing factorization or materialization  (Section~\ref{ssec:cost_model_FM}).

\para{Factorization vs.  materialization} 
The training process of an ML model requires complex arithmetic computations. Similar to  \emph{data warehousing} and \emph{virtual data integration}\footnote{We can classify the architectures of most traditional data integration systems as \emph{data warehousing} or \emph{virtual data integration} \cite{doan2012principles}. In data warehousing, the target table is  materialized and the materialization often requires an extract, transform, load (ETL) process. 
In a virtual data integration system, the target table is not materialized, and the user can pose queries against the target schema, also known as \emph{global schema} or \emph{mediated schema}. The user query needs to be rewritten according to the underlying source schemas, e.g., view-based query rewriting \cite{deutsch2006query,chirkova2012materialized}.},  these computations during model training can be conducted in a materialized or factorized manner.
Materialization requires joining the source tables and obtaining  the instances of the target table before exporting it for model training, as depicted in Figure~\ref{fig:motiv}. 
Another option is \emph{learning over factorized joins} \cite{10.1145/2882903.2882939}, also known as \emph{factorized learning} \cite{kumar2015learning}. 
Given an ML model and joinable tables of a database, factorized learning requires reformulating the ML model and pushing down the computation to each table. Compared to materialization, factorized learning does not affect model training accuracy but often helps to improve the training efficiency \cite{kumar2015learning, kumar_demonstration_2015, chen2017towards, alotaibi2021hadad, 10.1145/3299869.3319878, cheng_efficient_2021, khamis_acdc_2018, dsilva_aida_2018, dsilva_making_2019, 10.1145/2882903.2882939, schleich_layered_2019}. 
Notably, similar to traditional DI systems,  materialization is not possible in some cases due to privacy constraints and other reasons, which we address   in Section~\ref{sec:fl}. 
In this section, we focus on the performance implications of these two strategies.

\subsection{Computation challenge:  DI metadata for   factorization}
\label{ssec:LAR}
In the following, we explain how the data integration metadata can be used to generalize existing factorization techniques over silos (cf. Table~\ref{tbl:4cases}), and the new challenges. 
The existing factorization either tackles the model as a whole \cite{10.1145/2882903.2882939, kumar2015learning} or at the linear algebra level \cite{chen2017towards}, for linear or non-linear models \cite{kumar2015learning, kumar_demonstration_2015, chen2017towards, alotaibi2021hadad, 10.1145/3299869.3319878, cheng_efficient_2021, khamis_acdc_2018, dsilva_aida_2018, dsilva_making_2019}. To simplify the discussion, here we use the example of LA operator \emph{left matrix multiplication (LMM)} and its rewrite rule from \cite{chen2017towards}.


\para{New algebraic rewriting rules} Given a matrix $X$ with the size $c_T \times c_X$, the LMM of T and X is denoted as $TX$. 
For better understanding, we use   mapping/indicator matrices $M_k$/$I_k$ below, although we generate and utilize their compressed forms  $CM_k$ and $CI_k$ in practice.
Equations below present an example of transforming an existing LMM rewriting \cite{chen2017towards} with our proposed rule.

\vspace{-4mm}
    
\begin{gather}
 TX \rightarrow I_{1} (D_1 X[1:c_{S_1},]) + I_{2} (D_2 X[c_{S_1} +1 : c_T,]) \text{\cite{chen2017towards}}
\\ \nonumber \Downarrow
\\ TX \rightarrow \ I_{1} D_1 M_1^{T} X\  + \ ((I_{2} D_2  M_2^{T})\circ R_2 ) X \ \text{[Amalur]}
\vspace{-0.2cm} 
\end{gather} 

\noindent\emph{\circled{1} Local result generation.} We first compute    $I_{k} D_k M_k^{T}$ for each source table. In this step, to reduce computation overhead, we reorder the matrix multiplication sequence, similar to the join-order optimization in databases. 

\noindent\emph{\circled{2} Local result assembly.} The main task here is to detect and remove duplicate computations by applying the redundancy matrices. 
For instance, we continue with the running example.
Consider $D_1$ as the base table while $D_2$ is redundant.
To obtain the correct final LMM result, 
here we can perform a Hadamard Product $\circ$  (element-wise multiplication) between $I_{2} D_2  M_2^{T} $ and the redundancy matrix $R_2$. This way, we drop the redundant intermediate results indicated by the redundancy matrix $R_2$. 
Figure~\ref{fig:allMs}c shows the results of $T_1 X$ and $ (T_2\circ R_2) X$. It is easy to verify that their addition is the same as $TX$.

\para{DI metadata \& factorization} 
First, to compute the local LMM result,
in the above  rule (1) \cite{chen2017towards},   X is partitioned as $ X[1:c_{S_1}]$ and $ X[c_{S_1} +1 : c_T,]$ because the columns of $T$ are assumed to be two \emph{disjoint} sets from $D_1$ and $D_2$. 
To tackle the overlapping columns, in our modified rule (2), mapping matrix $M_k$ brings more flexibility in choosing the columns of $S_k$. 
Second, to compute the final result, 
in the original rule (1), two local LMM results (i.e., $D_1 X[1:c_{D_1},]$  and $D_2 X[c_{D_1}+1:c_T,]$) are simply  added up  via  indicator matrices $I_1$ and $I_2$. 
However, as we have shown, we need to handle redundancy when generalizing the LA factorization problem. 

\para{Challenges} Based on the process we described, we showed a simple example of how DI metadata can be used in ML factorization. Nonetheless, such processes might be less straightforward due to more complex schema  or row mappings, produced by the corresponding DI processes.
For example,  consider the cases where we have $1:n$ mappings among the schema attributes of the source tables and the one of the target table (e.g. \emph{fullname} mapping to \emph{first name} and \emph{last name}),  or the cases where source tables contain duplicated information (i.e., repeated entities) and require dedicated solutions. Embedding such DI metadata into factorization techniques is part of our future directions.

\subsection{Cost estimation challenge: to factorize or to materialize} 

\label{ssec:cost_model_FM}



\begin{figure}[tb]
\centering
\includegraphics[width=0.7\columnwidth]{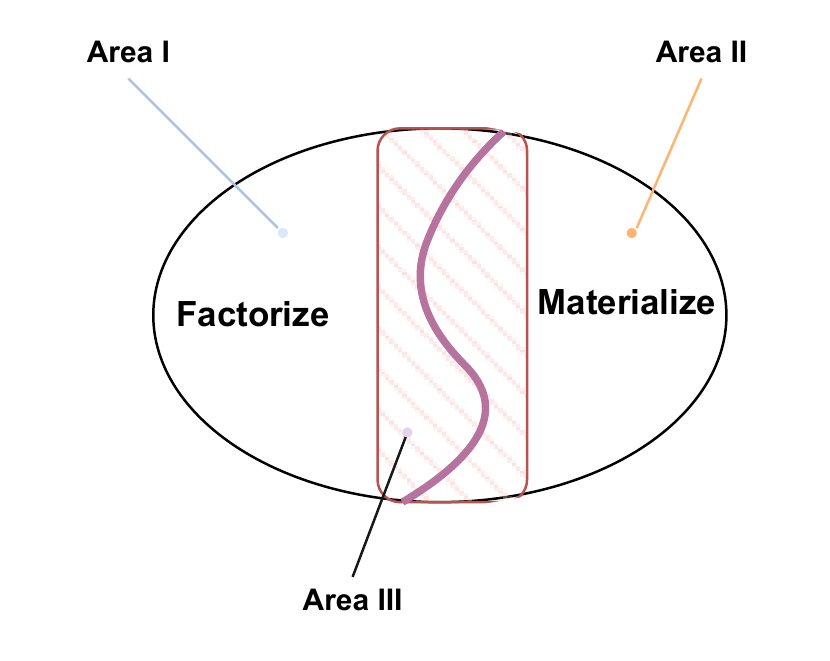}
 \vspace{-0.3cm}
\caption{An abstraction of different decision areas (factorize/materialize) and their boundaries}
\label{fig:ForM}
\vspace{-0.4cm}
\end{figure}

\para{Problem analysis} Factorization has been shown to be effective at increasing the efficiency of model training \cite{10.1145/2882903.2882939, kumar2015learning, kumar_demonstration_2015, chen2017towards, alotaibi2021hadad, 10.1145/3299869.3319878, cheng_efficient_2021, khamis_acdc_2018, dsilva_aida_2018, dsilva_making_2019, 10.1145/2882903.2882939, schleich_layered_2019}. 
However, the question of \emph{when} to factorize, is not fully answered. We illustrate the problem intuitively in Figure~\ref{fig:ForM}.
Let us assume that there exists a borderline (the curvy purple line), between the cases where factorization is faster and the cases where materialization is faster. Areas I and II cover the cases when it is easy to decide on factorization or materialization, respectively, while area III covers the harder cases. 
The state-of-the-art solution \cite{chen2017towards} only resolves the cases in Area I, missing many potential cases in Area III where factorization is faster. 

Essentially, cost estimation depends on four factors: the ML model, LA operators, hardware and underlying data. Given a model, its architecture and algebraic computations are fixed; it is known which  LA operators are affected by factorization and which are not \cite{chen2017towards}. 
To examine the relative speedup of factorization, we  mainly need to inspect the interactions between physical data transfers (e.g., network and memory bandwidth) \cite{DPP}, and data redundancy \cite{chen2017towards, 10.1145/2882903.2882939}.  
In general terms, if by joining the source tables we obtain a target table with  more instance redundancy than source tables, factorization may be faster than materialization. In the following, we explain why data silos bring more challenges and research opportunities.

\para{Cost estimation depends on data relationships} To differentiate area I from the rest in Figure~\ref{fig:ForM}, two heuristic rules are proposed based on tuple ratio and column ratio (i.e., feature ratio) between source tables and target tables \cite{chen2017towards, 10.1145/3299869.3319878, cheng_efficient_2021} . However, when we come to complex data integration scenarios, there are more parameters to consider. 
Before materializing the target table, among silos there are parameters relevant for the redundancy, source description (e.g., number of sources, number of columns and rows in each source, null value ratio per table), source correspondences (column matching and row matching between sources), etc. It is challenging to calculate the borderline in Figure~\ref{fig:ForM} and to design an accurate  cost model taking such data integration metadata into account.
From the preliminary experimental  results  
in Table~\ref{fig:amalur_performance}\footnote{Experiment setting:
$c_{S_1} =1$, $c_{S_2} = 100$; we set the values of $r_{S_1}$ as $\{10, 50, 100, 500, ..., 1000000, 5000000\}$, and 
$r_{S_2}$ as $0.2 \times r_{S_1}$, respectively. 
We tested ten scenarios in each of the four cases in Table~\ref{fig:amalur_performance}. We computed the percentage of times that the cost estimation procedures correctly predicted factorization.}, we observe that, by combining these parameters, in certain cases Amalur's cost estimation process can lead to better decisions than the state-of-the-art solution \emph{Morpheus} \cite{chen2017towards}.
In future work, we plan to cover a wide range of DI scenarios with more parameter combinations and rigorously test our approach. 

\para{DI metadata as cost model parameters and pruning rules}  Data integration metadata is relevant in two cases. First, as discussed above, some of DI metadata could be used as parameters of a cost model. The data integration principles could also potentially affect how these parameters should be combined.
Second, data integration is a topic with mature  logic-based theoretical frameworks. A natural question is: \emph{can we use these logic rules in cost estimation?}

\begin{table}[tb]
\centering
\resizebox{\columnwidth}{!}{
\begin{tabular}{ll|ll|}
\cline{3-4}
                                                         &     & \multicolumn{2}{c|}{\textbf{Redundancy in source tables}}                                                                                                                             \\ \cline{3-4} 
                                                         &     & \multicolumn{1}{c|}{\emph{Yes}}                                                                   & \multicolumn{1}{c|}{\emph{No}}                                                                    \\ \hline
\multicolumn{1}{|l|}{\multirow{3}{*}{\textbf{Redundancy in the target table}}} & \emph{Yes} & \multicolumn{1}{l|}{\begin{tabular}[c]{@{}l@{}}Morpheus: 70\%\\ Amalur: 70\%\end{tabular}} & \begin{tabular}[c]{@{}l@{}}Morpheus: 70\%\\ Amalur: 70\%\end{tabular} \\ \cline{2-4} 
\multicolumn{1}{|l|}{}                                   & \emph{No}  & \multicolumn{1}{l|}{\begin{tabular}[c]{@{}l@{}}Morpheus: 20\%\\ \textbf{Amalur: 80\%}\end{tabular}} & \begin{tabular}[c]{@{}l@{}}Morpheus: 30\%\\ \textbf{Amalur: 70\%}\end{tabular} \\ \hline
\end{tabular}
}
\caption{Percentage of correct factorization decisions of Amalur vs Morpheus \cite{chen2017towards}
}
\label{fig:amalur_performance}
\vspace{-0.4cm}
\end{table}
\vspace{-0.2cm}
\begin{exmp} 
	\label{exmp:tgd} 
Consider Example 2 in Table~\ref{tbl:4cases}.  
$m_1$ is a full tgd, i.e., $m_1$ does not contain existentially quantified variables.
All the attributes of target schema $T$ come from at least one source schema. 
In such a case, the number of columns in $T$ is less than or equal to the total number of columns in $S_1$ and $S_2$ participating in factorization. In the use cases of feature augmentation and VFL, the number of rows in $T$ is usually less than or equal to the total number of rows in $S_1$ and $S_2$. That is, the materialized target table $T$ does not contain more redundancy than the source tables. Thus, factorization will not bring performance improvement, which makes it a case in area II of Figure~\ref{fig:ForM}. 
In similar cases, we can make a straightforward decision on choosing materialization. 
A tgd is a first-order sentence, which can be easily implemented and evaluated, e.g., a datalog program. In future work, we plan to study how to utilize   such logical rules for cost estimation.
\end{exmp}
\vspace{-0.2cm}	
 
The above example is one of the simplest applications of mapping formalism in the context of cost estimation for factorization. There are more types of tgds describing more complicated dataset relationships, e.g., nested tgds \cite{kolaitis2020language}, and plain SO tgds \cite{arenas2013language}. The discussion could also be expanded to more types of metadata, e.g., expand the existing entity resolution approaches \cite{christophides2020overview} and come up with other pruning rules. In short, utilizing DI metadata in factorization still needs more effort from data integration theoreticians and practitioners.



\begin{myframe}{~Summary \& Opportunities~}
\emph{ -- Data integration metadata is useful for factorization over silos, and it requires much more research.}  
\\ \emph{ -- The more complicated dataset relationships call for new cost models to tell apart factorization from materialization.}
\end{myframe}

\section{Data integration and Federated Learning}
\label{sec:fl}

\subsection{Challenge I: Automate data transformation for FL}
\label{ssec:FL1}
In federated learning, a crucial prerequisite is establishing alignments among data silos, i.e., obtaining their column and row matching. This typically requires ML engineers to prepare a subset of local data by adding or removing feature and instance candidates from different data silos. It costs massive workforce or programming efforts to collect, prepare and transform data from the sources, which also involves tiresome re-engineering. 
With our proposed mapping and indicator matrices, the subsets of local data can be represented and embedded in the federated models, which has great potential to automate the whole process. In what follows, we explain our intuition with the vertical federated linear regression (FLR) algorithm from~\cite{yang@2019fml} and Example 2 in Table~\ref{tbl:4cases}.
The FLR training objective is:

\begin{align*}
\vspace{-0.3cm}	
			\begin{split} 
				\min\limits_{\Theta_A, \Theta_B} \sum\limits_{i} \left \| \Theta_{A} X_A^{(i)} + \Theta_{B} X_B^{(i)} - Y^{(i)}\right \|^2
			\end{split},
\vspace{-0.3cm}	
		\end{align*}
where $X_A$ and $X_B$ are feature spaces of $S_1$ and $S_2$ respectively, $Y$ is the label space of $S_1$, $\Theta_{A}$, and $\Theta_{B}$ are the local FLR model parameters of $S_1$ and $S_2$. $i$ denotes the row index of data instances in the matrix.

The performance of trained FLR models depends on the quality of $X_A$ and $X_B$, which are prepared before training and fixed during training. Refining the performance of FLR models typically requires regenerating $X_A$ and $X_B$. With our mapping and indicator matrices, we can integrate the generation of $X_A$ and $X_B$ into the FLR training as an end-to-end optimization procedure. By denoting $X_A$ as $I_1 D_1 M_1^T$ and $X_B$ as $I_2 D_2 M_2^T$, we rewrite the FLR objective as:

	\begin{align*}
			\begin{split} 
				\min_{\substack{\Theta_A, \Theta_B; \\ I_1,M_1, I_2,M_2}}  \sum\limits_{i} & \left \| \Theta_{A} (I_1 D_1 M_1^T)^{(i)}  + \Theta_{B} (I_2 D_2 M_2^T)^{(i)}  - Y^{(i)}\right \| 
			\end{split}
	\end{align*}
	
\para{Optimization and automation opportunities} The new FLR objective can be optimized by alternatively training $\Theta_A, \Theta_B$ and $I_1,M_1, I_2,M_2$. While $\Theta_A, \Theta_B$ can be trained by following the secure federated learning procedure~\cite{yang@2019fml}, efficiently and effectively training $ I_1,M_1, I_2,M_2$ are challenging problems. For one thing, the conventional centralized data selection approaches cannot be directly applied to the decentralized federated training because they impose heavy communication overheads. Therefore, communication-efficient data selection solutions are required to favor a fast search for optimal $ I_1, M_1, I_2, M_2$. For another, $ I_1, M_1, I_2, M_2$ contain metadata of different data sources, and therefore they should be trained in a privacy-preserving manner. These challenges open up new research directions that incorporate multiple disciplines involving data integration, federated learning, and cryptography.

\subsection{Challenge II: privacy-preserving DI+FL pipeline}
 
\label{ssec:FL2}
\para{Problem setting in existing FL frameworks} In   existing vertical federated learning frameworks for tabular data \cite{cheng2021secureboost, 10.1145/3514221.3526127, 10.1145/3448016.3457241, 10.1145/3459637.3482361}, the common assumption is that  entity resolution   and federated learning  are two isolated tasks, and entity resolution can be seen as a preprocessing step of federated learning. 
Moreover, they assume that the two data sources, party A and party B, do not have overlapping feature columns, as shown in the below example:
\\ Party A: $S_1(r,m, a)$
\\ Party B: $S_2(r,o)$, and the schema mapping is 
\\ $ m_1:\ \forall r, m, a,o \ (S_1(r,m, a) \wedge S_2(r,o)   \rightarrow T(m,a,o) )$  
\vspace{0.2cm}
\\ Source table $S_1$ of party A and source table $S_2$ of Party B, share a row id column $r$, which is not a feature. The row id matching is given to the VFL framework as inputs.  

However, such a problem setting is too ideal for real-life use cases. 
First, the results from an entity resolution approach often need to be verified by a human being. The separation between entity resolution   and federated learning means that a third party is needed. She  or he is trustworthy and will not leak information. Such a third party is not always available in a privacy-preserving use case. 
Second, as shown in Table~\ref{tbl:4cases}, the dataset relationships can be more complicated than the above example. The columns between silos could overlap, which requires a more sophisticated protocol to exchange gradients and model weights between different parties.
Thus, the natural question is: \emph{how to seamlessly apply the machine-generated ER results  to jointly train models while preserving  privacy?}   

\para{New challenges} The challenges of building an end-to-end entity resolution and federated learning pipeline are three-fold.
First, we need to define the representation for the machine-generated ER results, which are most likely approximate. Then the question is how to use such ER results in federated learning without affecting model accuracy significantly. 
Second, such a representation should not leak information about private data.
Intuitively, we can understand privacy preservation  as: each party will not learn new information about other parties during learning process, which they did not know before. The common techniques for privacy-preserving in federated learning and data integration   include homomorphic encryption \cite{fontaine2007survey, paillier1999public},  secret sharing \cite{shamir1979share, beimel2011secret} and differential privacy \cite{dwork2008differential}. In a system like Amalur where both entity resolution and federated learning are supported, we need to encrypt the data and also the data integration metadata. 
Third, there are new opportunities for time-wise efficiency improvement. It is well-known that encryption often brings tremendous computation overhead. Now besides the data, we have more to encrypt and decrypt, i.e., the metadata. 
The DI metadata is  generally smaller, compared to data instances. However, it is unclear how much overhead the encryption of DI metadata will bring, which requires further study. 

\begin{myframe}{~Summary \& Opportunities~}
\emph{ -- Data integration metadata is promising for improving ML model accuracy, and automating data transformation pipelines and federated learning.}  
\\ \emph{ -- More complicated dataset relationships bring more practical FL use cases, but also more challenges for encryption.} 
\end{myframe}

\section{Conclusion}
\label{sec:con}

In this work, we have explored the possibilities of bringing data integration and machine learning together. 
Towards this direction, we have proposed a novel data integration system Amalur, which supports data integration and machine learning over silos.
We have inspected the promising challenges of representing data integration metadata, and utilizing it for factorized learning and federated learning.
We envision this work as  one of the first steps towards
bridging the recent advances in machine learning with the well-studied traditional data integration field.



\section*{Acknowledgment}
This work  is co-funded  by the  European Union Horizon Programme call HORIZON-CL4-2022-DATA-01, under Grant Agreement No. 101093164 (ExtremeXP). 
\balance

\bibliographystyle{IEEEtran}
\bibliography{IEEEabrv,main}

\begin{thebibliography}{10}
\providecommand{\url}[1]{#1}
\csname url@samestyle\endcsname
\providecommand{\newblock}{\relax}
\providecommand{\bibinfo}[2]{#2}
\providecommand{\BIBentrySTDinterwordspacing}{\spaceskip=0pt\relax}
\providecommand{\BIBentryALTinterwordstretchfactor}{4}
\providecommand{\BIBentryALTinterwordspacing}{\spaceskip=\fontdimen2\font plus
\BIBentryALTinterwordstretchfactor\fontdimen3\font minus
  \fontdimen4\font\relax}
\providecommand{\BIBforeignlanguage}[2]{{%
\expandafter\ifx\csname l@#1\endcsname\relax
\typeout{** WARNING: IEEEtran.bst: No hyphenation pattern has been}%
\typeout{** loaded for the language `#1'. Using the pattern for}%
\typeout{** the default language instead.}%
\else
\language=\csname l@#1\endcsname
\fi
#2}}
\providecommand{\BIBdecl}{\relax}
\BIBdecl

\bibitem{bos2018prediction}
J.~M. Bos, G.~A. Kalkman, H.~Groenewoud, P.~M. van~den Bemt, P.~A. De~Smet,
  J.~E. Nagtegaal, A.~Wieringa, G.~J. van~der Wilt, and C.~Kramers,
  ``Prediction of clinically relevant adverse drug events in surgical
  patients,'' \emph{PloS one}, vol.~13, no.~8, p. e0201645, 2018.

\bibitem{hard2018federated}
A.~Hard, K.~Rao, R.~Mathews, S.~Ramaswamy, F.~Beaufays, S.~Augenstein,
  H.~Eichner, C.~Kiddon, and D.~Ramage, ``Federated learning for mobile
  keyboard prediction,'' \emph{arXiv preprint arXiv:1811.03604}, 2018.

\bibitem{doan2012principles}
A.~Doan, A.~Halevy, and Z.~Ives, \emph{{Principles of data integration}}.\hskip
  1em plus 0.5em minus 0.4em\relax Elsevier, 2012.

\bibitem{rahm2001survey}
E.~Rahm and P.~A. Bernstein, ``A survey of approaches to automatic schema
  matching,'' \emph{the VLDB Journal}, vol.~10, no.~4, pp. 334--350, 2001.

\bibitem{fagin2009clio}
R.~Fagin, L.~M. Haas, M.~Hern{\'{a}}ndez, R.~J. Miller, L.~Popa, and
  Y.~Velegrakis, ``{Clio: Schema mapping creation and data exchange},'' in
  \emph{ER}.\hskip 1em plus 0.5em minus 0.4em\relax Springer, 2009, pp.
  198--236.

\bibitem{brizan2006survey}
D.~G. Brizan and A.~U. Tansel, ``A. survey of entity resolution and record
  linkage methodologies,'' \emph{Communications of the IIMA}, vol.~6, no.~3,
  p.~5, 2006.

\bibitem{Makrynioti2019survey}
N.~Makrynioti and V.~Vassalos, ``{Declarative Data Analytics: a Survey},''
  \emph{TKDE}, p.~1, 2019.

\bibitem{zhou2020database}
X.~Zhou, C.~Chai, G.~Li, and J.~Sun, ``Database meets artificial intelligence:
  {A} survey,'' \emph{{IEEE} Trans. Knowl. Data Eng.}, vol.~34, no.~3, pp.
  1096--1116, 2022.

\bibitem{10.1007/978-3-030-35514-2_32}
M.~Schleich, D.~Olteanu, M.~Abo-Khamis, H.~Q. Ngo, and X.~Nguyen, ``{Learning
  Models over Relational Data: A Brief Tutorial},'' in \emph{Scalable
  Uncertainty Management}, N.~{Ben Amor}, B.~Quost, and M.~Theobald, Eds.\hskip
  1em plus 0.5em minus 0.4em\relax Cham: Springer International Publishing,
  2019, pp. 423--432.

\bibitem{DBLP:journals/tois/AlserafiARC20}
A.~Alserafi, A.~Abell{\'{o}}, O.~Romero, and T.~Calders, ``{Keeping the Data
  Lake in Form: Proximity Mining for Pre-Filtering Schema Matching},''
  \emph{{ACM} Trans. Inf. Syst.}, vol.~38, no.~3, pp. 26:1--26:30, 2020.

\bibitem{10.1145/3318464.3389742}
R.~Cappuzzo, P.~Papotti, and S.~Thirumuruganathan, ``{Creating Embeddings of
  Heterogeneous Relational Datasets for Data Integration Tasks},'' in
  \emph{Proceedings of the 2020 ACM SIGMOD International Conference on
  Management of Data}, 2020, pp. 1335--1349.

\bibitem{kopcke2010evaluation}
H.~K{\"o}pcke, A.~Thor, and E.~Rahm, ``Evaluation of entity resolution
  approaches on real-world match problems,'' \emph{Proceedings of the VLDB
  Endowment}, vol.~3, no. 1-2, pp. 484--493, 2010.

\bibitem{das2017falcon}
S.~Das, P.~S. GC, A.~Doan, J.~F. Naughton, G.~Krishnan, R.~Deep, E.~Arcaute,
  V.~Raghavendra, and Y.~Park, ``Falcon: Scaling up hands-off crowdsourced
  entity matching to build cloud services,'' in \emph{Proceedings of the 2017
  ACM International Conference on Management of Data}, 2017, pp. 1431--1446.

\bibitem{9338287}
Z.~Wang, B.~Sisman, H.~Wei, X.~L. Dong, and S.~Ji, ``Cordel: A contrastive deep
  learning approach for entity linkage,'' in \emph{2020 IEEE International
  Conference on Data Mining (ICDM)}, 2020, pp. 1322--1327.

\bibitem{krishnan2016activeclean}
S.~Krishnan, J.~Wang, E.~Wu, M.~J. Franklin, and K.~Goldberg, ``Activeclean:
  Interactive data cleaning for statistical modeling,'' \emph{Proceedings of
  the VLDB Endowment}, vol.~9, no.~12, pp. 948--959, 2016.

\bibitem{krishnan2017boostclean}
S.~Krishnan, M.~J. Franklin, K.~Goldberg, and E.~Wu, ``Boostclean: Automated
  error detection and repair for machine learning,'' \emph{arXiv preprint
  arXiv:1711.01299}, 2017.

\bibitem{9458702}
P.~Li, X.~Rao, J.~Blase, Y.~Zhang, X.~Chu, and C.~Zhang, ``Cleanml: A study for
  evaluating the impact of data cleaning on ml classification tasks,'' in
  \emph{2021 IEEE 37th International Conference on Data Engineering (ICDE)},
  2021, pp. 13--24.

\bibitem{dong2018data}
X.~L. Dong and T.~Rekatsinas, ``Data integration and machine learning: A
  natural synergy,'' in \emph{Proceedings of the 2018 international conference
  on management of data}, 2018, pp. 1645--1650.

\bibitem{DBLP:conf/vldb/HalevyRO06}
A.~Y. Halevy, A.~Rajaraman, and J.~J. Ordille, ``{Data Integration: The Teenage
  Years},'' in \emph{{VLDB}}.\hskip 1em plus 0.5em minus 0.4em\relax ACM Press,
  2006, pp. 9--16.

\bibitem{yang2019federated}
Q.~Yang, Y.~Liu, Y.~Cheng, Y.~Kang, T.~Chen, and H.~Yu, \emph{Federated
  Learning}.\hskip 1em plus 0.5em minus 0.4em\relax Morgan \& Claypool
  Publishers, 2019.

\bibitem{lenzerini2002data}
M.~Lenzerini, ``{Data integration: A theoretical perspective},'' in
  \emph{PODS}.\hskip 1em plus 0.5em minus 0.4em\relax ACM, 2002, pp. 233--246.

\bibitem{DBLP:conf/pods/GolshanHMT17}
B.~Golshan, A.~Y. Halevy, G.~A. Mihaila, and W.-C. Tan, ``{Data Integration:
  After the Teenage Years},'' in \emph{{PODS}}, 2017, pp. 101--106.

\bibitem{hai2016constance}
R.~Hai, S.~Geisler, and C.~Quix, ``{Constance: An Intelligent Data Lake
  System},'' in \emph{{SIGMOD}}.\hskip 1em plus 0.5em minus 0.4em\relax ACM,
  2016, pp. 2097--2100.

\bibitem{li2022metadata}
Z.~Li, R.~Hai, A.~Bozzon, and A.~Katsifodimos, ``Metadata representations for
  queryable {ML} model zoos,'' 2022.

\bibitem{voigt2017eu}
P.~Voigt and A.~Von~dem Bussche, ``The eu general data protection regulation
  (gdpr),'' \emph{A Practical Guide, 1st Ed., Cham: Springer International
  Publishing}, vol.~10, no. 3152676, pp. 10--5555, 2017.

\bibitem{quix2016metadata}
C.~Quix, R.~Hai, and I.~Vatov, ``{Metadata Extraction and Management in Data
  Lakes With GEMMS},'' \emph{CSIMQ}, no.~9, pp. 67--83, 2016.

\bibitem{chen2017towards}
L.~Chen, A.~Kumar, J.~Naughton, and J.~M. Patel, ``Towards linear algebra over
  normalized data,'' \emph{PVLDB}, vol.~10, no.~11, 2017.

\bibitem{10.1145/3299869.3319878}
S.~Li, L.~Chen, and A.~Kumar, ``{Enabling and Optimizing Non-Linear Feature
  Interactions in Factorized Linear Algebra},'' in \emph{Proceedings of the
  2019 International Conference on Management of Data}, 2019, pp. 1571--1588.

\bibitem{cheng_efficient_2021}
Z.~Cheng, N.~Koudas, Z.~Zhang, and X.~Yu, ``Efficient {Construction} of
  {Nonlinear} {Models} over {Normalized} {Data},'' Apr. 2021, pp. 1140--1151.

\bibitem{beeri1984proof}
C.~Beeri and M.~Y. Vardi, ``{A proof procedure for data dependencies},''
  \emph{JACM}, vol.~31, no.~4, pp. 718--741, 1984.

\bibitem{Fagin2009}
R.~Fagin, \emph{Tuple-Generating Dependencies}.\hskip 1em plus 0.5em minus
  0.4em\relax Boston, MA: Springer US, 2009, pp. 3201--3202.

\bibitem{chepurko2020arda}
N.~Chepurko, R.~Marcus, E.~Zgraggen, R.~C. Fernandez, T.~Kraska, and D.~Karger,
  ``Arda: automatic relational data augmentation for machine learning,''
  \emph{VLDB}, vol.~13, no.~9, pp. 1373--1387, 2020.

\bibitem{esmailoghli2021cocoa}
M.~Esmailoghli, J.-A. Quian{\'e}-Ruiz, and Z.~Abedjan, ``Cocoa: Correlation
  coefficient-aware data augmentation.'' in \emph{EDBT}, 2021, pp. 331--336.

\bibitem{kumar2016join}
A.~Kumar, J.~Naughton, J.~M. Patel, and X.~Zhu, ``{To join or not to join?
  thinking twice about joins before feature selection},'' in \emph{Proceedings
  of the 2016 International Conference on Management of Data}, 2016, pp.
  19--34.

\bibitem{yang@2019fml}
Q.~Yang, Y.~Liu, T.~Chen, and Y.~Tong, ``Federated machine learning: Concept
  and applications,'' \emph{ACM Trans. Intell. Syst. Technol.}, vol.~10, no.~2,
  jan 2019.

\bibitem{DBLP:conf/sigmod/ScannapiecoFBE07}
M.~Scannapieco, I.~Figotin, E.~Bertino, and A.~K. Elmagarmid, ``{Privacy
  preserving schema and data matching},'' in \emph{Proceedings of the {ACM}
  {SIGMOD} International Conference on Management of Data, Beijing, China, June
  12-14, 2007}.\hskip 1em plus 0.5em minus 0.4em\relax ACM, 2007, pp. 653--664.

\bibitem{xla}
A.~Sabne, ``{XLA} : Compiling machine learning for peak performance,'' 2020,
  {SIGMOD} workshop DEEM, industry keynote.

\bibitem{trt}
H.~Vanholder, ``Efficient inference with tensorrt,'' in \emph{GPU Technology
  Conference}, vol.~1, 2016, p.~2.

\bibitem{pytorch}
A.~Paszke, S.~Gross, F.~Massa, A.~Lerer, J.~Bradbury, G.~Chanan, T.~Killeen,
  Z.~Lin, N.~Gimelshein, L.~Antiga, A.~Desmaison, A.~Kopf, E.~Yang, Z.~DeVito,
  M.~Raison, A.~Tejani, S.~Chilamkurthy, B.~Steiner, L.~Fang, J.~Bai, and
  S.~Chintala, ``{PyTorch}: An imperative style, high-performance deep learning
  library,'' in \emph{Advances in Neural Information Processing Systems 32},
  2019, pp. 8024--8035.

\bibitem{tqp}
D.~He, S.~C. Nakandala, D.~Banda, R.~Sen, K.~Saur, K.~Park, C.~Curino,
  J.~Camacho{-}Rodr{\'{\i}}guez, K.~Karanasos, and M.~Interlandi, ``Query
  processing on tensor computation runtimes,'' \emph{Proc. {VLDB} Endow.},
  vol.~15, no.~11, pp. 2811--2825, 2022.

\bibitem{tcudb}
Y.-C. Hu, Y.~L. Li, and H.-W. Tseng, ``{TCUDB}: Accelerating database with
  tensor processors,'' in \emph{Proceedings of the 2022 International
  Conference on Management of Data}, 2022.

\bibitem{tensordb}
M.~Kim and K.~S. Candan, ``Tensordb: In-database tensor manipulation with
  tensor-relational query plans,'' in \emph{Proceedings of the 23rd ACM
  International Conference on Conference on Information and Knowledge
  Management}, 2014, pp. 2039--2041.

\bibitem{hummingbird}
D.~Koutsoukos, S.~Nakandala, K.~Karanasos, K.~Saur, G.~Alonso, and
  M.~Interlandi, ``Tensors: An abstraction for general data processing,''
  \emph{Proceedings of the VLDB Endowment}, vol.~14, no.~10, pp. 1797--1804,
  2021.

\bibitem{tpu}
N.~P. Jouppi, C.~Young, N.~Patil, D.~Patterson, G.~Agrawal, R.~Bajwa, S.~Bates,
  S.~Bhatia, N.~Boden, A.~Borchers \emph{et~al.}, ``In-datacenter performance
  analysis of a tensor processing unit,'' in \emph{Proceedings of the 44th
  annual international symposium on computer architecture}, 2017, pp. 1--12.

\bibitem{tensorcore}
\BIBentryALTinterwordspacing
NVIDIA, ``Nvidia tesla v100 gpu architecture,'' 2017. [Online]. Available:
  \url{http://images.nvidia.com/content/volta-architecture/
  pdf/volta-architecture-whitepaper.pdf}
\BIBentrySTDinterwordspacing

\bibitem{matrixparallel}
A.~R. Benson and G.~Ballard, ``A framework for practical parallel fast matrix
  multiplication,'' \emph{ACM SIGPLAN Notices}, vol.~50, no.~8, pp. 42--53,
  2015.

\bibitem{nature}
A.~Fawzi, M.~Balog, A.~Huang, T.~Hubert, B.~Romera-Paredes, M.~Barekatain,
  A.~Novikov, F.~J. R~Ruiz, J.~Schrittwieser, G.~Swirszcz \emph{et~al.},
  ``Discovering faster matrix multiplication algorithms with reinforcement
  learning,'' \emph{Nature}, vol. 610, no. 7930, pp. 47--53, 2022.

\bibitem{deutsch2006query}
A.~Deutsch, L.~Popa, and V.~Tannen, ``Query reformulation with constraints,''
  \emph{ACM SIGMOD Record}, vol.~35, no.~1, pp. 65--73, 2006.

\bibitem{chirkova2012materialized}
R.~Chirkova, J.~Yang \emph{et~al.}, ``Materialized views,'' \emph{Foundations
  and Trends{\textregistered} in Databases}, vol.~4, no.~4, pp. 295--405, 2012.

\bibitem{10.1145/2882903.2882939}
M.~Schleich, D.~Olteanu, and R.~Ciucanu, ``{Learning Linear Regression Models
  over Factorized Joins},'' in \emph{Proceedings of the 2016 International
  Conference on Management of Data}, 2016, pp. 3--18.

\bibitem{kumar2015learning}
A.~Kumar, J.~Naughton, and J.~M. Patel, ``{Learning generalized linear models
  over normalized data},'' in \emph{{SIGMOD}}, 2015, pp. 1969--1984.

\bibitem{kumar_demonstration_2015}
A.~Kumar, M.~Jalal, B.~Yan, J.~Naughton, and J.~M. Patel, ``Demonstration of
  {Santoku}: optimizing machine learning over normalized data,''
  \emph{Proceedings of the VLDB Endowment}, vol.~8, no.~12, pp. 1864--1867,
  Aug. 2015.

\bibitem{alotaibi2021hadad}
R.~Alotaibi, B.~Cautis, A.~Deutsch, and I.~Manolescu, ``Hadad: A lightweight
  approach for optimizing hybrid complex analytics queries,'' in \emph{SIGMOD},
  2021, pp. 23--35.

\bibitem{khamis_acdc_2018}
M.~A. Khamis, H.~Q. Ngo, X.~Nguyen, D.~Olteanu, and M.~Schleich,
  ``\BIBforeignlanguage{en}{{AC}/{DC}: {In}-{Database} {Learning}
  {Thunderstruck}},'' in \emph{\BIBforeignlanguage{en}{Proceedings of the
  {Second} {Workshop} on {Data} {Management} for {End}-{To}-{End} {Machine}
  {Learning}}}.\hskip 1em plus 0.5em minus 0.4em\relax Houston TX USA: ACM,
  Jun. 2018, pp. 1--10.

\bibitem{dsilva_aida_2018}
J.~V. D'silva, F.~De~Moor, and B.~Kemme, ``\BIBforeignlanguage{en}{{AIDA}:
  abstraction for advanced in-database analytics},''
  \emph{\BIBforeignlanguage{en}{Proceedings of the VLDB Endowment}}, vol.~11,
  no.~11, pp. 1400--1413, Jul. 2018.

\bibitem{dsilva_making_2019}
J.~V. D'silva, F.~{De Moor}, and B.~Kemme, ``Making an {RDBMS} data scientist
  friendly: Advanced in-database interactive analytics with visualization
  support,'' \emph{Proc. {VLDB} Endow.}, vol.~12, no.~12, pp. 1930--1933, 2019.

\bibitem{schleich_layered_2019}
M.~Schleich, D.~Olteanu, M.~Abo~Khamis, H.~Q. Ngo, and X.~Nguyen, ``A layered
  aggregate engine for analytics workloads,'' in \emph{Proceedings of the 2019
  International Conference on Management of Data}, 2019, pp. 1642--1659.

\bibitem{DPP}
M.~Zhao, N.~Agarwal, A.~Basant, B.~Gedik, S.~Pan, M.~Ozdal, R.~Komuravelli,
  J.~Pan, T.~Bao, H.~Lu, S.~Narayanan, J.~Langman, K.~Wilfong, H.~Rastogi,
  C.-J. Wu, C.~Kozyrakis, and P.~Pol, ``Understanding data storage and
  ingestion for large-scale deep recommendation model training,'' in
  \emph{Proceedings of the 49th Annual International Symposium on Computer
  Architecture}.\hskip 1em plus 0.5em minus 0.4em\relax {ACM}, jun 2022.

\bibitem{kolaitis2020language}
P.~G. Kolaitis, R.~Pichler, E.~Sallinger, and V.~Savenkov, ``On the language of
  nested tuple generating dependencies,'' \emph{TODS}, vol.~45, no.~2, pp.
  1--59, 2020.

\bibitem{arenas2013language}
M.~Arenas, J.~P{\'{e}}rez, J.~Reutter, and C.~Riveros, ``{The language of plain
  SO-tgds: Composition, inversion and structural properties},'' \emph{Journal
  of Computer and System Sciences}, vol.~79, no.~6, pp. 763--784, 2013.

\bibitem{christophides2020overview}
V.~Christophides, V.~Efthymiou, T.~Palpanas, G.~Papadakis, and K.~Stefanidis,
  ``An overview of end-to-end entity resolution for big data,'' \emph{ACM
  Computing Surveys (CSUR)}, vol.~53, no.~6, pp. 1--42, 2020.

\bibitem{cheng2021secureboost}
K.~Cheng, T.~Fan, Y.~Jin, Y.~Liu, T.~Chen, D.~Papadopoulos, and Q.~Yang,
  ``{Secureboost: A lossless federated learning framework},'' \emph{IEEE
  Intelligent Systems}, vol.~36, no.~6, pp. 87--98, 2021.

\bibitem{10.1145/3514221.3526127}
F.~Fu, H.~Xue, Y.~Cheng, Y.~Tao, and B.~Cui, ``{BlindFL}: Vertical federated
  machine learning without peeking into your data,'' in \emph{{SIGMOD} '22:
  International Conference on Management of Data, Philadelphia, PA, USA, June
  12 - 17, 2022}.\hskip 1em plus 0.5em minus 0.4em\relax {ACM}, 2022, pp.
  1316--1330.

\bibitem{10.1145/3448016.3457241}
F.~Fu, Y.~Shao, L.~Yu, J.~Jiang, H.~Xue, Y.~Tao, and B.~Cui, \emph{{VF2Boost:
  Very Fast Vertical Federated Gradient Boosting for Cross-Enterprise
  Learning}}.\hskip 1em plus 0.5em minus 0.4em\relax Association for Computing
  Machinery, 2021, pp. 563--576.

\bibitem{10.1145/3459637.3482361}
W.~Fang, D.~Zhao, J.~Tan, C.~Chen, C.~Yu, L.~Wang, L.~Wang, J.~Zhou, and
  B.~Zhang, ``Large-scale secure {XGB} for vertical federated learning,'' in
  \emph{Proceedings of the 30th ACM International Conference on Information \&
  Knowledge Management}, 2021, pp. 443--452.

\bibitem{fontaine2007survey}
C.~Fontaine and F.~Galand, ``A survey of homomorphic encryption for
  nonspecialists,'' \emph{EURASIP Journal on Information Security}, vol. 2007,
  pp. 1--10, 2007.

\bibitem{paillier1999public}
P.~Paillier, ``Public-key cryptosystems based on composite degree residuosity
  classes,'' in \emph{International conference on the theory and applications
  of cryptographic techniques}.\hskip 1em plus 0.5em minus 0.4em\relax
  Springer, 1999, pp. 223--238.

\bibitem{shamir1979share}
A.~Shamir, ``How to share a secret,'' \emph{Communications of the ACM},
  vol.~22, no.~11, pp. 612--613, 1979.

\bibitem{beimel2011secret}
A.~Beimel, ``Secret-sharing schemes: A survey,'' in \emph{International
  conference on coding and cryptology}.\hskip 1em plus 0.5em minus 0.4em\relax
  Springer, 2011, pp. 11--46.

\bibitem{dwork2008differential}
C.~Dwork, ``Differential privacy: A survey of results,'' in \emph{International
  conference on theory and applications of models of computation}.\hskip 1em
  plus 0.5em minus 0.4em\relax Springer, 2008, pp. 1--19.

\end{thebibliography}

\end{document}